# SLiMFast: Guaranteed Results for Data Fusion and Source Reliability

Manas Joglekar, Theodoros Rekatsinas, Hector Garcia-Molina, Aditya Parameswaran[†], Christopher Ré

Stanford University and [†] University of Illinois, Urbana-Champaign

## ABSTRACT

We focus on data fusion, i.e., the problem of unifying conflicting data from data sources into a single representation by estimating the source accuracies. We propose SLiMFast, a framework that expresses data fusion as a statistical learning problem over discriminative probabilistic models, which in many cases correspond to logistic regression. In contrast to previous approaches that use complex generative models, discriminative models make fewer distributional assumptions over data sources and allow us to obtain rigorous theoretical guarantees. Furthermore, we show how SLiMFast enables incorporating domain knowledge into data fusion, yielding accuracy improvements of up to 50% over state-of-the-art baselines. Building upon our theoretical results, we design an optimizer that obviates the need for users to manually select an algorithm for learning SLiMFast's parameters. We validate our optimizer on multiple real-world datasets and show that it can accurately predict the learning algorithm that yields the best data fusion results.

## 1. INTRODUCTION

Integrating information from multiple data sources is crucial for maximizing the value extracted from data. Different data sources can provide information about the same *object*, e.g., a real-world entity or event, but this information can be inconsistent. That is, data provided by different sources may be in conflict. Thus, *data fusion*—the task of resolving conflicts across sources by estimating their trustworthiness—has emerged as a key element of many data integration pipelines [14].

Interacting with collaborators from a medical school—who are currently engaged in extracting information from scientific articles to populate a structured data repository—we find that using data fusion can be challenging for users despite the numerous approaches proposed in the literature [9, 15, 22, 23, 29, 39]. We demonstrate some challenges in current data fusion methods by example, and argue that expressing data fusion as a statistical learning problem over discriminative probabilistic models leads to methods which solve data fusion more accurately than existing approaches, and come with rigorous theoretical guarantees.

We use the application of our collaborators as an example to demonstrate challenges that users face due to limitations of current data fusion approaches, and thereby motivate our approach.

**Example 1.** *A genomics expert wants to extract* gene mutations associated with genetic diseases *from a corpus of 22 million PubMed articles. The expert's goal is to use this data to diagnose patients with Mendelian disorders. To this end, the expert uses information extraction techniques over scientific articles to collect (gene, disease, associated) triples where the field "associated" takes on values in $\{true, false\}$. An article can be thus viewed as providing*

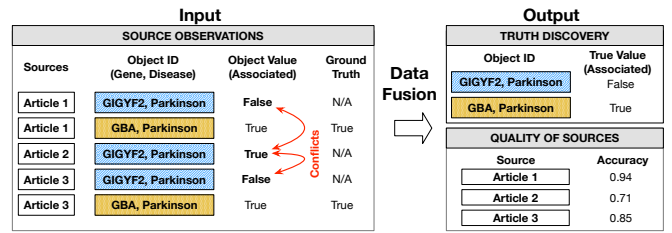

**Figure 1: A schematic diagram of data fusion when extracting gene-disease associations from scientific articles.**

*a set of triples that make claims about gene-disease associations. In many cases, the extracted triples may contain conflicts. For instance, the sentences "Variation in GIGYF2 is not associated with Parkinson disease"[1] and "These data strongly support GIGYF2 as a PARK11 gene with a causal role in familial Parkinson Disease"[2] will yield conflicting extractions. The expert wants to use data fusion to resolve such conflicts and obtain only gene-disease associations most likely to be correct according to the literature.*

For exposition, we view articles as data sources. Other notions of sources, such as research labs or journals, can be naturally captured in the framework described below. The extracted (gene, disease, associated) tuples from each article can be viewed as observations for a collection of objects with a *gene-disease* pair being the object's id and *associated* the object's value. Data fusion resolves conflicts by estimating the trustworthiness of data sources and uses that to estimate the true value of each object. The trustworthiness of a data source is quantified via the notion of *source accuracy*, i.e., the probability that an observation made by the source is correct. Figure 1 shows an overview of the input and output of data fusion. The input is a set of source observations and the output consists of the estimated source accuracies and the true values of objects. In certain cases, limited ground truth on the correctness of source observations may be available and can be used to obtain an initial estimate on the accuracy of data sources.

**The Need for Guarantees.** Continuing with our example, the domain expert wants to use the extracted data for patient diagnosis. Due to the critical nature of the diagnosis application, formal guarantees on the correctness of the objects in the final knowledge base are required. As a result, data fusion should also provide formal guarantees that the returned gene-disease associations are correct within a certain margin of error.

One approach to obtain guarantees is having access to a siz-able amount of labeled (ground truth) data and use that to estimate source accuracies. However, obtaining large volumes of labeled data can be prohibitively expensive due to the monetary cost associated with human annotators. Hence, most existing data fusion approaches assume no ground truth data. This forces them to rely on procedures such as *expectation maximization* (EM) to estimate the accuracy of data sources. These procedures come with few theoretical guarantees on their convergence and may yield suboptimal solutions [38]. The above raises the key technical question we seek to address, i.e., *how much ground truth is actually needed to obtain high-quality results with formal guarantees?* We show that under certain conditions, which match our target applications, only a surprisingly small amount of ground truth data is sufficient to obtain data fusion techniques that simultaneously (i) identify the correct values of objects with high-confidence, and (ii) obtain low-error (in many cases less than 2%) estimates of source accuracies. Low-error accuracy estimates of sources are crucial in intelligence applications [32, 34] and can also help users minimize the monetary cost of data acquisition by purchasing only accurate data sources [12]. Finally, in certain cases, the presence of ground truth eliminates the need for iterative procedures such as EM and allows us to use highly-efficient techniques for solving data fusion, thus, leading to overall scalable methods for free.

---

[1] http://www.neurology.org/content/72/22/1886.abstract
[2] http://www.ncbi.nlm.nih.gov/pubmed/18358451

**The Need for Domain Knowledge.** The expert believes that findings in some articles may not be trustworthy altogether and wants to use additional external knowledge about the sources themselves to estimate their accuracy better and further improve the quality of the data fusion result. Information such as the citation count or publication year can be informative of the accuracy of an article's claims: the expert believes that highly-cited or more recent articles are more trustworthy. More fine-grained information such as the experimental design of a study can also be informative. For instance, the genomics expert trusts results of gene-knockout studies, but is skeptical about genome-wide association studies (GWAS).

Current data fusion methods do not consider domain knowledge but only use the conflicting observations across sources to estimate their accuracy. Moreover, existing methods rely on complex models that are not easily extensible to incorporate *domain-specific features*. In this paper, we demonstrate how domain knowledge can be integrated in data fusion in the form of domain-specific features that are indicative of a data source having high or low accuracy.

Apart from the expert in genomics, we also interviewed experts in applied micro-economic theory, and a consumer electronics company, all engaged in extracting facts from the scientific literature to support analytic applications. The requirements outlined above were unanimously identified as important across all domains. These requirements were also outlined in a recent survey by Li et al. [25] as open problems for data fusion. We show how our data fusion approach can address these open problems.

*Our Approach.* To address the above challenges, we introduce SLiMFast, a framework that uses discriminative probabilistic models to perform data fusion. At its simplest form, SLiMFast's model corresponds to *logistic regression*. This formulation separates data fusion into two tasks: (i) performing *statistical learning* to compute the parameters of the graphical model—used to estimate the accuracy of data sources—and (ii) performing *probabilistic inference* to predict the true values of objects. Using probabilistic models for data fusion is not new [6, 9, 15, 29, 39]. However, SLiMFast comes with several advancements over previous work:

(1) SLiMFast is the *first data fusion approach to combine cross-source conflicts with domain-specific features*—integrated as addi-

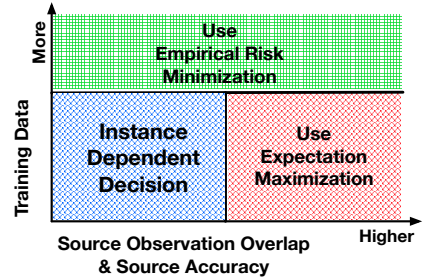

**Figure 2: The tradeoff space SLiMFast's optimizer captures to select the best algorithm for learning the source accuracies.**

tional variables in SLiMFast's probabilistic model—to solve data fusion more accurately. We show that, for real-world applications, combining these two signals can yield accuracy improvements of up to 50% at identifying the true value of objects and can lead up to $10\times$ lower error in source accuracy estimates.

(2) While not obvious at first glance, we show that many existing data fusion approaches correspond to a special case of SLiMFast which uses a logistic regression model. In turn, this allows us to use standard techniques from statistical learning theory [28] to provide *strong guarantees for data fusion when limited ground truth data is available*; in our experimental evaluation we show that in certain cases as few as 10 training examples are sufficient to obtain an accuracy higher than 0.9 when estimating the true value of objets.

(3) To learn the parameters in SLiMFast, we can either use expectation maximization (EM) or empirical risk minimization (ERM) if ground truth is available. The performance of the two algorithms depends on different aspects of the input data. ERM relies on ground truth data, while EM is affected by the overlap across source observations and the average accuracy of sources. This introduces a tradeoff (Figure 2) between the two algorithms in terms of which can estimate the source accuracies better. To automate the decision between EM and ERM, we design an *optimizer that determines which algorithm will lead to more accurate data fusion results*. Internally, our optimizer uses the notion of *units of information*: assuming that one labeled example for the true value of an object has one unit of information, we design a statistical model that estimates the equivalent units of information in source observations used by the EM algorithm. We evaluate our optimizer on multiple real-world datasets with varying properties, and show that in almost all cases it selects the best performing algorithm correctly.

*Summary of Contributions.* We propose SLiMFast, a framework that expresses data fusion as a statistical learning problem over discriminative probabilistic models. We show how SLiMFast answers several open problems in data fusion, outlined in a recent survey [25]. In Section 2, we review data fusion. Then, in Section 3, we describe SLiMFast and show that many existing data fusion methods are captured by SLiMFast, thus, obviating the need for *model selection*. In Section 4, we provide a series of guarantees for the output of SLiMFast. We present a series of *theoretical guarantees* for data fusion and design an optimizer that automatically selects the best algorithm for learning SLiMFast's parameters. In Section 5, we evaluate SLiMFast and its optimizer on multiple real-world datasets. We also demonstrate how domain-specific features unlock additional functionalities, such as obtaining insights on the accuracy of data sources and solving the problem of *source reliability initialization*.

## 2. PRELIMINARIES

We review data fusion in the context of data integration and introduce the main terminology and notation used in the paper. In SLiMFast, we consider integrating data from a set of data sources $S$ that follow a common representation (e.g., they have the same schema in the case of relational data). Sources provide data describing different objects (e.g., named entities or real-world events). As in existing approaches, SLiMFast assumes that different objects described in the data sources are already identified and aligned across data sources. We point the reader to a survey by Bleiholder and Naumann [6] that describes how common schema matching and duplicate detection techniques can be used to address these two problems. We use $O$ to denote the set of distinct objects that sources in $S$ describe. We use the genomics example from Section 1 to demonstrate this.

**Example 2.** *In the genomics application, data sources $S$ correspond to scientific articles. After information extraction techniques are applied, a collection of tuples describing objects that correspond to gene-disease pairs, e.g., (GIGYF2, Parkinson), is extracted from each article. Each object is described by the attribute "associated" that takes values in {true, false} and indicates if the gene and disease associated with the object are truly associated or not according to the article. The set of all distinct gene-disease pairs from all available articles corresponds to the set of objects $O$.*

After alignment and duplicate detection, multiple, possibly inconsistent observations of the same objects are found across data sources. That is, the values assigned to an attribute of a real-world object by different sources may not agree. Informally, the goal of data fusion is to combine conflicting source observations for the same object into a single representation while inconsistencies across data sources are resolved.

**Example 3.** *In the genomics application (Figure 1), three articles provide information for object (GIGYF2, Parkinson). Two of them state that gene GIGYF2 is not associated with Parkinson, i.e., the attribute "associated" for object (GIGYF2, Parkinson) takes the value "false". In contrast to the first two articles, the third one that gene GIGYF2 is associated with Parkinson (i.e., the attribute "associated" takes the value "true").*

We focus on inconsistencies for a single object-attribute. Extending to multiple attributes is straightforward. Given a source $s \in S$ and an object $o \in O$, we denote $v_{o,s}$ the value that source $s$ assigns to the attribute of object $o$. We refer to each $v_{o,s}$ as a *source observation* and use $\Omega$ to denote the set of all source observations for all objects. Furthermore, we assume that, for the attribute under consideration, each object $o \in O$ has a true *latent* value $v_o^*$. Given the observations $\Omega$ from a set of sources $S$ for a set of objects $O$, the goal of data fusion is to estimate the latent true values $v_o^*$ for all objects in $O$ and output the estimates to the user. We use $v_o$ to denote the estimated true value of an object $o \in O$. Similarly to existing data fusion methods [9, 25, 40], we consider setups that follow *single-truth semantics*, i.e., there is only one correct value for each object $o$ and at least one source provides it. This assumption is related to closed-world semantics. Nonetheless, the models introduced in the remainder of the paper can support open-world semantics, i.e., allow for the true value of objects to not be reported by any source. This can be modeled by allowing variables $v_o^*$ to take a wildcard value corresponding to the unknown truth. For simplicity of presentation we do not study such extensions here.

In fusing data from multiple sources, one can distinguish between a variety of strategies. Simple strategies that estimate the true object values by assigning $v_o$ to the most often occurring value or the average value reported by different sources are widely adopted in data integration [6]. However, strategies that reason about the trustworthiness of data sources to resolve inconsistencies across observations have been shown to be more accurate in estimating the true values of objects [24]. Intuitively, instead of treating source observations in a uniform manner, the aforementioned approaches consider values reported for an object by more trustworthy sources to be more probable. We focus on such strategies.

To measure the trustworthiness of a data source SLiMFast uses the notion of *accuracy*. The true accuracy of a data source $s \in S$, denoted $A_s^*$, is defined as the probability that the information provided by $s$ for an object is correct [9, 39]. For modeling purposes, the accuracy of a data source is assumed to be the same across all objects, thus, the probability that an observation $v_{o,s}$ from a source $s \in S$ for an object $o \in O$ is correct is given by $P(v_{o,s} = v_o^*) = A_s^*$. This assumption is typical for data fusion and can be easily relaxed by allowing a source to have multiple accuracy parameters for different object classes [10, 17, 23].

Given the above probabilistic semantics, data fusion in SLiMFast is expressed as a statistical learning problem whose goal is to infer a function, parameterized by the unknown source accuracies $A_s^*$, that maps the input source observations $\Omega$ to output true value estimates $v_o$ for each object $o \in O$. Since the true source accuracies are unknown, SLiMFast needs to estimate the unknown accuracy of each source $s \in S$, denoted by $A_s$, and use that to estimate the unknown true values $v_o^*$ for objects in $O$. Similar probabilistic formulations are adopted by many data fusion frameworks [9, 15, 29, 39]. We follow probabilistic semantics as they promote interpretability. In the next section, we describe the components of SLiMFast in detail and compare it with existing data fusion methods.

## 3. THE SLiMFast FRAMEWORK

An overview of SLiMFast is shown in Figure 3. The core input to SLiMFast is a collection of source observations $\Omega$. Users also have the option to provide a set of labeled ground truth data, denoted $G$, corresponding to the true values for a subset of objects. Ground truth data (usually limited) is commonly used in data fusion to obtain initial estimates of the accuracy of data sources [12, 22, 40] before using iterative procedures to obtain the final estimates of source accuracies and the latent true values of objects.

**Example 4.** *In the genomics application (Figure 1), we have ground truth data that gene GBA is associated with Parkinson. This provides partial evidence that Article 1 and Article 3 are trustworthy since they state that GBA is truly associated with Parkinson.*

Finally, in SLiMFast, users have the option to specify a set of domain-specific features they deem to be informative of the accuracy of data sources. We use $K$ to denote the set of domain-specific features and $f_{s,k}$ to denote the value a data source $s \in S$ takes for a feature $k \in K$. We denote the collection of all $f_{s,k}$ values as $F$.

**Example 5.** *Examples of features characterizing the sources in the genomics application are shown in Figure 3. Since data sources correspond to scientific articles, examples of domain-specific features provided include the number of citations and the publication year of articles. For instance, source A1 is characterized by the features "publication year = 2009" and "citations = 34" while source A2 by "publication year = 2008" and "citations = 128".*

As a next step, SLiMFast compiles the input source observations $\Omega$, ground truth $G$, if provided, and the feature values $F$ to a probabilistic graphical model, and casts data fusion as a learning and inference problem over that model. Before learning and inference are performed, SLiMFast's internal optimizer analyzes (i) the structure of the graphical model obtained by the compilation step, and (ii) the available ground truth to determine the best learning algorithm to be used for estimating the accuracy of data sources. Finally, learning and inference are performed to solve data fusion. In addition to the traditional output of data fusion, SLiMFast leverages the presence of domain-specific features to provide additional functionalities to users, such as explanations as to which features are the most informative of source accuracies (Section 5.3.1). Next, we discuss the role of domain-specific features in data fusion, introduce SLiM-Fast's probabilistic graphical model, and describe SLiMFast's connections to existing data fusion models.

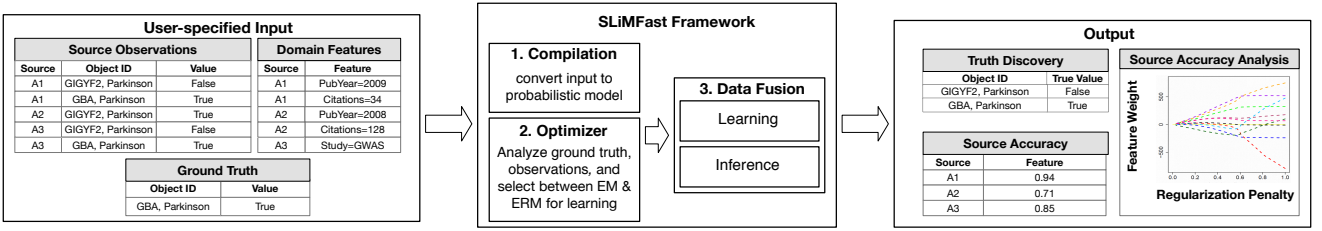

Figure 3: An overview of SLiMFast's core components. The provided source observations and domain-specific features are compiled into a probabilistic model used to solve data fusion via statistical learning and probabilistic inference.

### 3.1 The Role of Domain-Specific Features

Existing data fusion methods rely on conflicting or overlapping source observations to estimate the accuracy of data sources and to estimate the true values of objects. In addition to source observations, SLiMFast can also leverage user-specified domain-specific features as an additional signal to estimate the accuracy of sources. Combining these two signals has been identified as a promising direction for data fusion in prior work [11]. Indeed, our analysis (Section 4.2) shows that in large part due to domain features SLiMFast can provably obtain low-error estimates of the source accuracies and can determine the true values of objects accurately using a limited amount of ground truth. Thus, the presence of domain-specific features makes the use of ground truth in data fusion practical.

Obtaining the values of domain-specific features for data sources requires significantly lower effort than obtaining ground truth. This is because one can use automated techniques to collect informative metadata for data sources. To name a few examples, consider different types of data sources ranging from web pages to even crowd workers, e.g., workers from Amazon Mechanical Turk[3], and scientific articles. For web pages, it is easy to collect traffic statistics, such as the number of daily visitors, PageRank, and the page's bounce rate from third party portals. Such statistics can be informative of the web page's accuracy. For crowd workers, features such as the number of tasks performed or the average time required per task can be indicative of the accuracy of a worker's answers. Finally, for scientific articles, the publication venue, author, publication year and readability ease can be important. To promote flexibility across different domains, SLiMFast allows users to specify the features most relevant to their task.

### 3.2 SLiMFast's Data Fusion Model

SLiMFast expresses data fusion as a learning and inference problem over a probabilistic graphical model. For each object $o \in O$, SLiMFast introduces a latent random variable $T_o$ modeling the object's unknown true value $v_o^*$. Each source observation $v_{o,s}$ is mod-eled as an observed random variable $V_{o,s}$. We denote $D_o$ the set of distinct values that sources in $S$ assign to object $o$. To determine the estimated true value $v_o$ of an object, SLiMFast computes the posterior probability $P(T_o = d|\Omega)$ for all values $d \in D_o$, and assigns $v_o$ the value that maximizes the probability of variable $T_o$.

---

[3]https://www.mturk.com

SLiMFast follows a *discriminative* approach, i.e., it models the posterior $P(T_o|\Omega)$ directly considering the observations reported by sources in $S$ for object $o \in O$. For simplicity's sake, we consider that SLiMFast uses a logistic regression model over the source observations for object $o$. More elaborate models can be expressed in SLiMFast as we discuss in Section 3.3. We have:

$$P(T_o = d|\Omega) = \frac{1}{Z} \exp \sum_{(o,s) \in \Omega} \sigma_s \mathbb{1}_{v_{o,s}=d} \quad (1)$$

where $Z = \sum_{d \in D_o} \exp \sum_{(o,s) \in \Omega} \sigma_s \mathbb{1}_{v_{o,s}=d}$ is a normalization constant, parameter $\sigma_s$ denotes the trustworthiness score of source $s$ providing observation $v_{o,s}$, and $\mathbb{1}$ corresponds to the 0,1-indicator function. As in the mixture of expert models [19], $\sigma_s$ can be defined as the log odds that the observation provided by source $s \in S$ agrees with the unknown true value $v_o^*$ of object $o$:

$$\sigma_s = \log(\frac{P(v_{o,s} = v_o^*)}{1 - P(v_{o,s} = v_o^*)}) = \log(\frac{A_s^*}{1 - A_s^*}) \quad (2)$$

SLiMFast estimates the unknown accuracy of each data source via a logistic function model that is parameterized by the domain-specific features $K$ and an additional source-indicator feature for each source in $S$. For the estimated accuracy $A_s$ of data source $s \in S$ we have:

$$A_s = 1/(1 + \exp(-w_s - \sum_{k \in K} w_k f_{s,k})) \quad (3)$$

Model parameters $\langle w_k \rangle_{k \in K}$ capture the importance of domain-specific features for determining the accuracy of data sources. Parameters $\langle w_s \rangle_{s \in S}$ offer SLiMFast the flexibility to capture the heterogeneity of data sources and enable SLiMFast to recover existing data fusion models when no domain-specific features are specified. Combining Equations 1-3 gives us SLiMFast's final model:

$$P(T_o = d|\Omega; w) =$$
$$= \frac{1}{Z} \exp(\sum_{(o,s) \in \Omega} (w_s + \sum_{k \in K} w_k f_{s,k}) \mathbb{1}_{v_{o,s}=d}) \quad (4)$$

with $w = (\langle w_s \rangle_{s \in S}, \langle w_k \rangle_{k \in K})$.

*Compilation.* SLiMFast uses a *factor graph* representation to encode the above logistic regression model. Any declarative factor graph framework (e.g., Alchemy [1], DeepDive [2, 36], PSL [5]) can by used. Declarative factor graph frameworks allow one to easily expand SLiMFast's logistic regression model with additional features. For example, in Appendix D, we extend SLiMFast's model to capture pairwise correlations across copying data sources.

*Data Fusion with SLiMFast.* To solve data fusion SLiMFast needs to (i) learn the parameters $w$ of the logistic regression model in Equation 4 by optimizing the likelihood $\ell(w) = \log P(T|\Omega; w)$ where $T$ corresponds to the set of all variables $T_o$, and (ii) infer the maximum a posteriori (MAP) assignments to variables $T_o$.

When sufficient ground truth is available—we formalize this in Section 4—SLiMFast uses *empirical risk minimization* (ERM) to compute the parameters of its logistic regression model [27]. ERM sets SLiMFast's parameters $w$ in Equation 4 to values that maximize the likelihood of the object values provided in the ground truth. The optimization objective of ERM corresponds to the likelihood $\ell$ taken over the observed variables $T_o$ in the ground truth data. As no latent variables are involved in this step, the optimization objective is convex and efficient methods such as stochastic gradient descent (SGD) can be used to learn $w$. Afterwards, probabilistic inference is used to estimate the value $v_o$ of objects not present in the ground truth data. Variables $v_o$ are assigned to the maximum a posteriori (MAP) estimates of variables $T_o$.

If ground truth is limited or not available, SLiMFast uses *expectation maximization* (EM) to compute the parameters $w$ that maximize the likelihood of the source observations $\Omega$. EM estimates $w$ and $v_o$ iteratively by alternating between two steps: (i) the expectation step (E-step), where given an assignment to parameters $w$ the estimated true values of objects $v_o$ are assigned to the MAP estimates of variables $T_o$, and (ii) the maximization step (M-step), where given an assignment to variables $v_o$ the model parameters $w$ are estimated via their maximum likelihood values. When EM is used, variables $T_o$ associated with ground truth data correspond to observed random variables in the compiled factor graph. The value of the remaining latent variables is estimated using the above iterative procedure. This corresponds to a typical semi-supervised learning scenario. For EM, optimizing likelihood $\ell$ corresponds to a non-convex objective as EM jointly learns the parameters $w$ and the distribution of $T_o$ that maximizes $\ell$. The above iterative procedure can be expensive and may converge to local optima.

After the compilation phase is over, SLiMFast's optimizer determines which of the two algorithms (ERM or EM) should be used and then learning and inference are performed within the framework used to express SLiMFast's probabilistic model. We used the DeepDive framework [2, 36]. Probabilistic inference is performed via Gibbs sampling. All ERM, EM and Gibbs sampling are implemented over DeepDive's sampler [41].

## 3.3 SLiMFast and Existing Fusion Methods

Several existing approaches express data fusion as a learning and inference problem over probabilistic graphical models [9, 15, 29, 39]. The core difference between SLiMFast and previous methods is that SLiMFast uses a *discriminative* probabilistic model while existing data fusion methods use *generative* probabilistic models. As a result, SLiMFast does not make strong distributional assumptions on how the source observations are generated but estimates the conditional probability of the unknown true values of objects directly given the source observations. As we show in Section 5, the lack of strong distributional assumptions allows SLiMFast to solve data fusion more accurately.

In particular, certain data fusion models [9, 43] rely on Naive Bayes and assume that source observations are conditionally independent. On the other hand, SLiMFast uses logistic regression—Naive Bayes' discriminative equivalent [28]—which does not assume that observations are conditionally independent. When the dependent variables (in data fusion the latent true values of objects) are Boolean, then Naive Bayes and logistic regression are the same [27]. The latter implies that in certain cases, data fusion approaches designed around Naive Bayes can be expressed in SLiMFast. Finally, as we discuss in the next section, discriminative models allow us to obtain formal guarantees for data fusion by adapting standard tools from statistical learning theory.

The fact that we build SLiMFast over a declarative factor graph framework, gives us with the flexibility to express many of the existing data fusion methods in it. In general, one can follow the *discriminative relaxation* procedure introduced by Patel et al. [31] to convert data fusion methods that use sophisticated generative models to their discriminative counterparts, which, in turn, can be represented as a factor graph. While simple, this realization, combined with SLiMFast's optimizer, obviates the need for users to select a data fusion model for different data fusion tasks, thus, shedding light to the problem of model selection introduced by Li et al. [25].

## 4. DATA FUSION WITH GUARANTEES

We describe how SLiMFast yields data fusion solutions that come with rigorous error guarantees. In SLiMFast, we must choose between ERM and EM to solve data fusion. Below, we describe the key factors that determine the quality of data fusion results obtained by SLiMFast for these two algorithms. We then provide formal error bounds on SLiMFast's output estimates and build upon those to design an optimizer that analyzes SLiMFast's input and automatically selects which of the ERM and EM to use so as to maximize the quality of the returned data fusion solution.

### 4.1 Overview of SLiMFast's Guarantees

Empirical risk minimization and expectation maximization were described in Section 3.2. When ERM is used, the main factor that determines the quality of SLiMFast's is (i) the amount of ground truth. For EM, the main factors are (ii) the average accuracy of data sources, and (iii) the density of source observations, i.e., the average fraction of data sources providing observations for an object. To illustrate the impact of these factors we use a simple example:

**Example 6.** *We use a synthetic dataset with observations from 1,000 sources for 1,000 objects. We consider independent sources for simplicity. We measure how accurately SLiMFast estimates the true values of objects when ERM and EM are used.[4] We vary the average source accuracy in $[0.5, 0.8]$, the percentage of training data in $[.1\%, 60\%]$, and the density in $[0.005, 0.02]$. The results are shown in Figure 4. ERM is indeed affected only by the amount of ground truth and remains stable as the other two factors vary. The opposite is observed for EM.*

Motivated by Example 6, we study SLiMFast's probabilistic model to obtain guarantees for SLiMFast's data fusion solutions. We provide error bounds on the estimated true values of objects as well as the estimated accuracy of data sources. Our results support the experimental evidence from Example 6. We now provide a summary of our results and state them in detail in Section 4.2:

- When ground truth is available and ERM is used, standard tools from learning theory give us that the error for both the estimated object values and the estimated source accuracies is proportional to $\sqrt{\frac{|K|}{|G|}}$ where $|G|$ is the number of ground truth examples and $|K|$ the number of features in SLiMFast.
- When no ground truth is available and EM is used, we show that the error in estimated source accuracies is bounded by $\tilde{O}(\frac{1}{|S|\delta} + \sqrt{\frac{|K|}{|S||O|p}})$ where $|K|$ is the number of features in SLiMFast, $p$ is the probability of a source providing an observation for an object, thus, determining density, and $\delta \in (0, 0.5]$ is a parameter such that $A_s^* \geq 0.5 + \delta$ for each source $s \in S$.[5] This is a novel bound that does not rely on existing learning theory results.[6]

---
[4]The methodology and metrics are the same as in Section 5. EM and ERM correspond to Sources-EM and Sources-ERM.

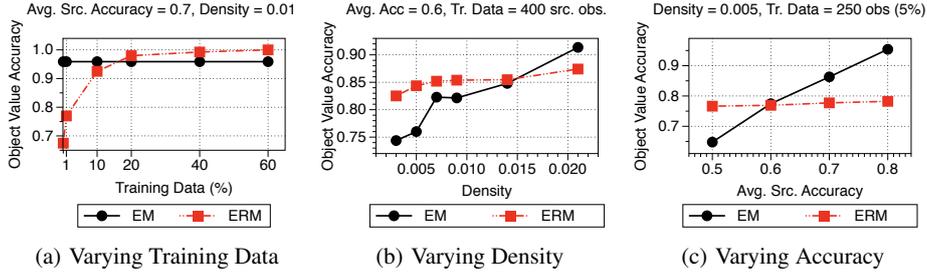

(a) Varying Training Data  (b) Varying Density  (c) Varying Accuracy

Figure 4: EM versus ERM as we vary (a) the amount of ground truth, (b) the density, and (c) the average accuracy of data sources. ERM is less sensitive to the fusion instance characteristics. EM performs better for more accurate sources and more sense instances.

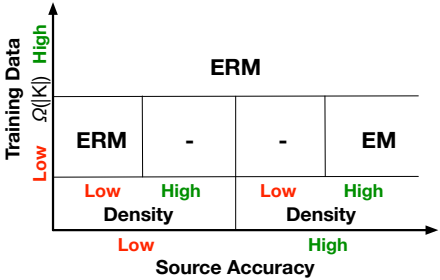

Figure 5: The tradeoff space for ERM and EM with respect to properties of a data fusion instance. For certain parts of the space, we report the algorithm that leads to more accurate data fusion results. Dash indicates that the best algorithm varies.

Our theoretical analysis reveals a tradeoff in the quality of solutions obtained by ERM and EM as the ground truth, average source accuracy, and density of a given data fusion instance vary. This tradeoff is evident in Figure 4. For sufficient ground truth ERM outperforms EM, but when ground truth is limited (as expected in real applications) the best algorithm varies. For fixed ground truth, EM can outperform ERM for dense instances (Figure 4(b)) or instances where the average source accuracy is high (Figure 4(c)). Figure 5 shows the tradeoff space. In summary, determining the most accurate algorithm is not trivial, thus, we develop an information theoretic optimizer for comparing the performance of EM and ERM. SLiMFast's optimizer is described in Section 4.3.

### 4.2 SLiMFast's Theoretical Foundations

We present the results summarized above in detail. Our analysis uses results and tools from learning theory, such as Rademacher complexity. A background discussion on those and the proofs for all theorems can be found in the Appendix of the paper.

---
[5]Notation $\tilde{O}$ subsumes logarithmic factors.
[6]While we do not obtain a bound for object values, previous work [16, 42] has shown that under strong generative assumptions such a bound can be obtained. Nevertheless, these assumptions are hard to evaluate in practice.

#### 4.2.1 Guarantees with Ground Truth Data

We use standard generalization bounds based on Rademacher complexity [26] to obtain guarantees on the estimated true values of SLiMFast's output. Given a parameter assignment $w$ for SLiMFast and a set of source-observations $\Omega$, SLiMFast computes the posterior probability for each variable $T_o$ using Equation 4. We have the following theorem bounding the risk of SLiMFast's model:

**Theorem 1** ([26]). *Let $w$ be a parameter assignment to SLiMFast. Fix distribution $\mathcal{D}$ to be the true distribution from the object values and source observations are generated. Fix parameter $\delta \in (0, 1)$. Let $L(w)$ be the expected log-loss under distribution $D$ when inferring the unknown values of objects using SLiMFast parameterized via $w$. Moreover, let $G$ correspond to a random sample of ground truth data on object values given as input to SLiMFast and $L_G(w)$ be the expected log-loss with respect to that ground truth sample. Then with probability $\geq 1 - \delta$ over the draw of $G$:*

$$L(w) - L_G(w) \leq O(\sqrt{|K|/|G|} \log(|G|))$$

This theorem states that the log-loss over all objects is not much larger than the log-loss over ground truth $G$. Thus, if we used ERM to learn the model parameters $w$ by optimizing over the log-loss $L_G(w)$ and if $\sqrt{|K|/|G|}$ is small, we will infer the correct true values of all objects with high-probability.

We now turn our attention to the estimated accuracy of data sources. We state the way we measure SLiMFast's performance. SLiMFast's model is parameterized by a parameter vector $w$ of length $|K|$. We assume that source-object pairs $(s, o)$ (and a corresponding true object value $v_o^*$ and observation $v_{s,o}$) are drawn from some unknown distribution $\mathcal{D}$. Let $\mathcal{D}_S$ denote the probability distribution over sources obtained from $\mathcal{D}$. Let $L(w)$ denote the *loss* of a model with weight vector $w$ which quantifies how bad its source accuracy estimates are with respect to distribution $\mathcal{D}$. We use again use the log-loss for this purpose:

**Definition 7** (Accuracy Estimate Loss). *The loss of a weight vector $w$, denoted $L_w$, is defined as:*

$$L_w = E_{(s,o) \sim \mathcal{D}}[L_w(s, o)]$$

*where*

$$L_w(s, o) = \mathbb{1}_{V_{o,s} = v_o^*} \log(A_s(w)) + \mathbb{1}_{V_{o,s} \neq v_o^*} \log(1 - A_s(w))$$

*and $\mathbb{1}$ is the indicator function, and $A_s$ is estimated by Equation 3.*

Loss $L_w$ can also be written as:

$$L_w = E_{s \sim \mathcal{D}_S}[A_s^* \log(A_s(w)) + (1 - A_s^*) \log(1 - A_s(w))]$$

where $A_s^*$ denotes the true unknown accuracy of a data source. This quantity is minimized when $A_s(w) = A_s^*$ for each $s$, and the difference from its optimal value is the expected Kullback-Leibler (KL) divergence between the true accuracy $A_s^*$ and the predicted accuracy $A_s(w)$ if the two accuracies are viewed as Bernoulli random variables. We can bound the accuracy estimation error as:

**Theorem 2.** *Fix distribution $\mathcal{D}$ and a parameter $\delta \in (0, 1)$. Let $G$ be a sample of training examples from distribution $\mathcal{D}$. Let*

$$w^* = \mathrm{argmin}_w L_w \text{ and } w = \mathrm{argmin}_w \sum_{(s,o) \in G} L_w(s, o)$$

*Then with probability $\geq 1 - \delta$ over the draw of $G$,*

$$L_w - L_{w^*} \leq O(\sqrt{|K|/|G|} \log(|G|))$$

This theorem also follows from results on Rademacher complexity. While similar to generalization bounds, here, the loss $L$ is parameterized by $w$. Our bound states that the estimation error for source accuracies is proportional to $\sqrt{|K|}$. This suggests that adding extra, potentially uninformative features to our model can worsen our source accuracy estimates. A user has no way of knowing which features are informative and which are not. Fortunately, it is possible for a user to add many features, without incurring a large increase in accuracy estimation error, as long as we regularize our model appropriately. $L_1$-regularization is a well-known method to induce sparsity in the solution [4], i.e., to yield only a small number of non-zero values in the parameter vector $w$. If the user provides a large number of features, of which only a few are predictive of source accuracies, then using $L_1$-regularization will obtain a vector $w$ that only assigns non-zero weights to the predictive features. Moreover, the error in the accuracy estimate is proportional to the square root of the number of predictive features, rather than the square root of the total number of features. Formally, if $k$ of the $|K|$ features get non-zero weight, then we have:

$$L_w - L_{w^*} \leq O(\sqrt{k \log(|K|)/|G|} \log(|G|))$$

### 4.2.2 Guarantees without Ground Truth

We now assume no ground truth and rely only on source observations for estimating the source accuracies. To simplify our theoretical analysis, we assume that:

- The true accuracy of each source $A_s^*$ takes values in $[0.5 + \delta/2, 1 - \delta/2]$ for some $0 < \delta \leq 0.5$.
- Sources have *uniform selectivities*, i.e., for each source-object pair $(s, o)$, the probability that source $s$ provides an observation for $o$ is $p$. We also assume that $p \geq 2/|S|$, i.e., in expectation, at least two sources give observations for an object. The expected number of observations is $|S||O|p$.
- We consider a set of $K$ features that are predictive of the accuracies of data sources.

**Theorem 3.** *If all above conditions hold we can obtain an estimate $A_s$ for the accuracy of source $s$ such that*

$$\frac{1}{|S|} \sum_{s \in S} D_{KL}(A_s \| A_s^*) \leq O\left(\frac{\log |O|}{|S|\delta} + \sqrt{\frac{|K|}{|O||S|p} \frac{\log^2(|O||S|)}{\delta}}\right)$$

*where $D_{KL}$ stands for the KL divergence.*

This theorem offers two important insights on unsupervised data fusion methods: their error depends on two properties of a fusion instance (i) the accuracy of sources, and (ii) the overlap across source observations. The higher source accuracies are (i.e., the higher $\delta$ is) the lower the estimation error will be. The same holds for the source overlap (controlled via $p$). The aforementioned assumptions were only used to simplify our analysis and are not inherent to SLiMFast. We only leverage the above insights to design a general model that allows SLiMFast to identify when EM can be used to solve data fusion accurately. In Section 5 we show that this model is accurate even when certain of the modeling assumptions do not hold (e.g., when most sources' accuracy is lower than 0.5).

### 4.3 SLiMFast's Optimizer

Considering the main steps performed by ERM and EM, described in Section 3.2, we see that while ERM uses the object values provided as ground truth to estimate parameters $w$ in SLiMFast, EM uses the estimated object values output from its E-step. Based on this observation, we can compare the performance of ERM and EM if we compare the information in the output of the E-step for EM with the information in the ground truth used in ERM.

*Comparing EM with ERM.* We consider the information in ground truth data: Given an object $o$ and its estimated true value $v_o$, we define a Boolean random variable $C_o$ taking the value "true" if $v_o$ obtains the correct value for $o$ and the value "false" otherwise. If we have no access to ground truth data the maximum value of $C_o$'s entropy is $H(C_o) = 1$. Given ground truth information on $v_o$ the entropy of $C_o$ becomes $H(C_o) = 0$, thus, we gain 1-unit of information. If we have $m$ sources providing observations for object $o$, the total information gain is $m$-units for this object. Summing units over all objects in ground truth, we obtain the total units of information gain. We apply a similar procedure to EM. We use an example to illustrate the intuition behind our approach:

**Example 8.** *Consider* 10 *sources providing binary observations for an object $o$ whose true value is unknown. We assume that all sources have the same accuracy* 0.7 *and assume majority vote is used to resolve conflicts. Since, the number of distinct values assigned by sources to object $o$ is two, majority vote will retrieve the correct value for $o$ only when more than 5 sources provide the correct value. The probability $p_e$ of this event occurring is given as a function of the CDF of a Binomial distribution:*

$$p_e = 1 - \sum_{i=0}^{5} \binom{10}{i} 0.7^i (1 - 0.7)^{10-i} = 0.8497$$

*We now consider the random variable $C_o$ from above. Given $p_e$ the probability that variable $C_o$ becomes true is $p_e$ and we have $H(C_o) = -p_e \log_2(p_e) - (1 - p_e) \log_2(1 - p_e) = 0.611$. Thus, after applying the majority vote model during the E-step, object $o$ contributes* 0.389-*units of information. Multiplying that with* 10 *we have that object $o$ contributes a total of* 3.89-*units.*

Generalizing the above example, we use the following model to estimate the information we gain after the E-step of EM: For an object $o$, let $m$ be the number of sources providing observations for it, and $|D_o|$ the number of distinct values assigned to $o$ by these sources. Our optimizer makes the assumption that all sources have the same accuracy $A$ and conflicts are resolved via majority vote. This model is used only in our optimizer to make a quick decision between EM and ERM as we discuss below. The probability that this model will obtain the correct value for an object is measured via the CDF of a Binomial distribution parameterized by $A$, $m$, and $m/|D_o|$. To estimate the total number of information units, we iterate over all objets. Algorithm 1 shows this procedure.

*Estimating the Average Source Accuracy.* Algorithm 1 requires the average accuracy of sources as input. To estimate this

**Algorithm 1:** EMUnits: Estimating Information Units for EM

**Input**: Objects $O$, Source Observations $\Omega$, Average Source Accuracy $A$
**Output**: EM Information Units
$totalUnits = 0$;
**for** *each object in $O$* **do**
    $m = \#$ of sources with observations for $o$;
    $|D_o| = \#$ of distinct values assigned to $o$ by sources;
    $p_e = 1 - \sum_{i=0}^{\lfloor m/|D_o| \rfloor} \binom{m}{i} A^i (1-A)^{m-i}$;
    **if** $p_e \geq 0.5$ **then**
        $totalUnits \mathrel{+}= 1 - p_e \log_2(p_e) - (1 - p_e) \log_2(1 - p_e)$;
**return** $totalUnits$;

unknown quantity, we rely on matrix completion. Given the set of source observations $\Omega$, we define $X$ to be an $|S| \times |S|$ matrix capturing the *agreement rate* of data sources. Given a data source $s$ let $O_s$ denote the set of objects for which $s$ provides an observation. We define entry $X_{i,j}$ for sources $s_i$ and $s_j$ as $X_{i,j} = \frac{1}{|O_{s_i} \cap O_{s_j}|} \sum_{o \in O_{s_i} \cap O_{s_j}} \mathbb{1}_{v_{o,s_i} = v_{o,s_j}} - \mathbb{1}_{v_{o,s_i} \neq v_{o,s_j}}$. We assume that all data sources have the same accuracy $A$ and that they are not adversarial, i.e., $A > 0.5$. Given $A$, the expected agreement rate between sources $s_i$ and $s_j$ is $A^2 + (1-A)^2 - 2A(1-A) = (2A-1)^2$. Let $\mu = 2A - 1$. The expected agreement rate between sources $s_i$ and $s_j$ is $E[X_{i,j}] = \mu^2$. Therefore, we estimate $\mu$ as $\hat{\mu} = \arg\min \frac{1}{2}\|X - \mu^2\|_2$. This optimization problem corresponds to a matrix completion problem. Setting the derivative of the optimization objective to zero gives as the closed form solution $\hat{\mu} = \sqrt{\frac{\sum_{i,j} X_{i,j}}{|S|^2 - |S|}}$. Given $\hat{\mu}$ the average accuracy of data sources $A$ is $A = (\hat{\mu} + 1)/2$. This setup can be extended to a different accuracy per source via a more general matrix completion problem. Matrix completion comes with optimality guarantees [7] and variants of SGD can be used to solve it efficiently [35].

**Algorithm 2:** SLiMFast's Optimizer

**Input**: Objects $O$, Source Observations $\Omega$, Source Features $K$, Ground Truth $G$, Threshold $\tau$
**Output**: Learning Algorithm
**if** $\sqrt{|K|/|G|} \log(|G|) < \tau$ **then**
    return ERM;
$totalERMUnits = |G|$;
Estimate Average Source Accuracy $A$ ;
$totalEMUnits = \text{EMUnits}(O,\Omega,A)$;
**if** *totalERMUnits < totalEMUnits* **then**
    return EM;
**else**
    return ERM;

*Overall Algorithm.* SLiMFast's optimizer uses Algorithm 2 to decide between ERM and EM. First, the algorithm examines if the bound in Equation 2 is below a threshold $\tau$. If so, then it always chooses the ERM algorithm. Otherwise, it compares the amount of ground truth with the estimated EM units (Algorithm 1) to decide between EM and ERM. In Section 5 we evaluate our optimizer on different real world datasets and show that it selects the best algorithm almost every time. Our optimizer runs in 2% of the total time required to solve data fusion.

**Table 1: Parameters of the data used for evaluation. For genomics the true average accuracy of data sources cannot be estimated reliably due to the sparsity of the dataset.**

| Parameter | Stocks | Demos | Crowd | Genomics |
|---|---|---|---|---|
| # Sources | 34 | 522 | 102 | 2750 |
| # Objects | 907 | 3105 | 992 | 571 |
| Available GrdTruth | 100% | 100% | 100% | 100% |
| # Observations | 30763 | 27736 | 19840 | 3052 |
| # Domain Features | 7 | 7 | 4 | 4 |
| # Feature Values | 70 | 341 | 171 | 16358 |
| Avg. Src. Acc. | < 0.5 | 0.604 | 0.540 | - |
| Avg. Obsrvs per Obj. | 33.9 | 15.703 | 20 | 5.345 |
| Avg. Obsrvs per Src. | 904.79 | 53.13 | 194.51 | 1.11 |

*Discussion.* When the number of labeled data is large, one can estimate the accuracy of sources as the empirical fraction of erroneous observations per source. Standard error guarantees apply here. Nonetheless, SLiMFast still uses ERM, as the above empirical estimator corresponds to Naive Bayes and the conditional independence assumption may not apply in practice (see Section 5).

## 5. EXPERIMENTAL EVALUATION

We compare SLiMFast against state-of-the-art data fusion techniques on four diverse real-world datasets. We show that SLiMFast yields accuracy improvements of up to 50% over state-of-the-art baselines for estimating the true values of objects and 2× to 10× lower error estimates for source accuracies.

The main points we seek to validate are: (i) how much training data is needed to obtain high-quality data fusion models, (ii) what is the impact of domain-specific features on data fusion, and (iii) how effective is SLiMFast's optimizer in selecting between EM and ERM for learning the parameters of SLiMFast's probabilistic model. Finally, we investigate how SLiMFast's output can be used to provide insights on source accuracies and study how more complex fusion methods can be expressed in SLiMFast.

### 5.1 Experiment Setup

We describe the datasets, metrics, and experimental settings used to validate SLiMFast against competing data fusion methods.

*Datasets.* We use four datasets, one from the finance domain, one from the intelligence domain, a crowdsourcing dataset, and a genomics dataset corresponding to typical data fusion scenarios. For the first two datasets we seek to integrate information from web pages on real-world objects including stocks and demonstration-events. In the third dataset the goal is to integrate answers provided by crowd-workers for a popular sentiment analysis task. The fourth dataset corresponds to the Genomics application from Section 1. Table 1 shows statistics for these datasets. All datasets follow the single-truth semantics, i.e., objects have one correct value and at least one source provides that.

*Stocks.* This is a popular data fusion dataset [24] where sources provide information on the volume of stocks, i.e., the total number of shares that trade hands from sellers to buyers, for July 2011. The dataset contains information on multiple stock-attributes. We focus on stock volumes as they exhibit the most conflicts across sources. We removed sources that are no longer active and for which no traffic statistics could be obtained (see discussion on domain-specific features below), and removed highly-accurate sources (e.g., NAS-DAQ) used to obtain ground truth data.

*Demonstrations.* This dataset contains reports of demonstrations in Africa from January, 2015 to April, 2015 from GDELT [21], a cat-

alogue with extractions of real-world events from online news articles. The same demonstration can be reported by multiple sources but the corresponding GDELT entries can contain conflicts (e.g., report different dates or lat-long coordinates) due to extraction errors. Each extraction is treated as an object taking values in {true, false} and our goal is to determine the correct extractions. Sources correspond to online news domains. GDELT data can be particularly challenging for data integration and data cleaning tasks [3, 33] due to the heterogeneity of data sources and noisy extractions. Ground truth was obtained using the ACLED dataset[7]. ACLED is a human-curated database of demonstrations in Africa. To generate ground truth data, we mapped each GDELT entry to an ACLED entry considering the date and location of the two entries. Any demonstration not in ACLED is assumed to be false.

*Crowd.* This is the "weather sentiment" dataset from Crowdflower [8]. The dataset contains crowd evaluations for the sentiment of weather-related tweets corresponding to *positive*, *negative*, *neutral*, and *not weather related*. It contains 1,000 tweets and contributions from 20 workers per tweet. Our goal is to detect the true sentiment for each tweet. Ground truth evaluations are provided with the raw data.

*Genomics.* This dataset was obtained from the Genetic Association Database (GAD). [9] The dataset contains gene-disease associations reported in the scientific literature. Each association is annotated by human experts as *positive* or *negative* if the authors claim a positive or a negative association respectively. Ground truth was obtained from human-curated datasets available at DisGeNet. [10] From the full dataset we only considered gene-phenotype associations that have conflicting observations from at least two sources.

*Domain-specific Features.* For the first two datasets, sources correspond to web-domains. For each domain we obtained traffic statistics from Alexa.com: (i) global rank, (ii) country rank, (iii), bounce rate, (iv) daily page views per visitor, (v) daily time on site, (vi) search visits, and (vii) total sites linking in. All metrics take numeric values and are discretized to get Boolean features. We found that discretization does not affect SLiMFast's performance significantly. For Crowd, features are (i) the channel of each worker, i.e., the particular market used to hire the worker, (ii) the country and (iii) city of the worker, and (iv) the fraction of tweets labeled by the worker. For Genomics, we extracted the (i) journal, (ii) citation count, (iii) publication year, and (iv) author list from PubMed.

*Methods.* We compare three variations of SLiMFast against baselines. First, we focus on methods using discriminative models:

- **SLiMFast-ERM, SLiMFast-EM, SLiMFast**: The first two always use ERM and EM respectively. The last uses SLiMFast's optimizer to select between EM and ERM.
- **Sources-ERM**: Same as SLiMFast but without domain-specific features. ERM is always used.
- **Sources-EM**: The same as Sources-ERM but EM is always used. This approach can be viewed as the discriminative equivalent of the generative model proposed Zhao et al. [43].

We also consider methods that use generative models:

- **Counts**: This corresponds to Naive Bayes. Source accuracies are estimated as the fraction of times a source provides the correct value for an object in ground truth.

- **ACCU**: This is the Bayesian data fusion method introduced by Dong et al. [9]. We do not consider source copying.

Finally, we compare SLiMFast against two state-of-the-art iterative optimization data fusion methods [22, 39]. Iterative optimization methods do not rely on probabilistic semantics but measure the trustworthiness of data sources via a reliability weight:

- **CATD**: This data fusion method was introduced by Li et al. [22] and extends source reliability scores with confidence intervals to account for sparsity in source observations.
- **SSTF**: This data fusion method by Yin et al. [40] leverages semi-supervised graph learning to exploit the presence of ground truth data.

*Different Methods and Ground Truth.* ERM-based methods use ground truth to learn the model parameters. For EM-based methods, ground truth is used as evidence in the factor graph and EM is executed until convergence, thus, corresponding to a semi-supervised approach. For ACCU and CATD, any available ground truth is used to initialize the source accuracy estimates, as suggested in [9] and [22]. Both algorithms are executed until convergence.

*Evaluation Methodology.* All datasets come with ground truth for all objects. In our experiments, we focus on small amounts of training data up to 20%. We vary the percentage of training data in {0.1, 1, 5, 10, 20}. The splits are generated randomly, thus, for each fraction of training data we run each method five times and report the average performance. To measure performance we use:

- *Accuracy for True Object Values:* the fraction of objects for which the data fusion method identifies the correct true value over the total number of objects. The accuracy is computed with respect to objects present in the testing data.
- *Error for Estimated Sources Accuracies:* a weighted-average of the absolute estimation error across all source accuracies. For each source we take the absolute error between its estimated accuracy and its true accuracy. The true accuracies are computed using all ground truth data. Errors are weighted by the number of observations per source to penalize wrong estimates for sources that provide many observations. A similar weighting scheme is used by existing approaches [22, 23].

*Implementation Details.* SLiMFast, Sources-ERM and Sources-EM, are implemented over DeepDive v0.7. EM and ERM are implemented on top of DeepDive's sampler using SGD to optimize the likelihood objective. All other methods are implemented in Python. All experiments were executed on a machine with four CPUs (each CPU is a 12-core 2.40 GHz Xeon E5-4657L), 1TB RAM, running Ubuntu 12.04. While all methods run in memory, their footprint is significantly smaller than the available resources.

## 5.2 Experimental Results

We provide a comparison to competing data fusion methods on the quality of the data fusion output. We demonstrate that in most cases SLiMFast outperforms state-of-the-art data fusion methods. We also evaluate our optimizer and show that it enables SLiMFast to correctly choose between EM and ERM (in all but one cases) so that it obtains the best output for data fusion. A comparison of the running time of different methods is provided in Appendix C.

### 5.2.1 Identifying the True Value of Objects

We evaluate how accurately different data fusion methods estimate the true object values, and report the relative difference between SLiMFast and competing approaches. The results are shown in Table 2. For SLiMFast, we report the results obtained when using our optimizer with the threshold $\tau$ in Algorithm 2 set to 0.1. The effect of $\tau$ on our optimizer is studied in Section 5.2.3. As shown in Table 2 Panel B, on average, SLiMFast outperforms other data fusion methods significantly with relative accuracy differences of more than 10% in many cases.

---
[7] http://www.acleddata.com
[8] http://www.crowdflower.com/data-for-everyone
[9] https://geneticassociationdb.nih.gov/
[10] http://disgenet.org/

**Table 2: Accuracy for predicting the true object values with varying training data (TD). Bold indicates the best performing method. Sources-ERM(-EM) is reported as S-ERM(-EM).**

Panel A: Accuracy for different types of data fusion methods.

| Dataset | TD (%) | Discriminative | | | Generative | | Iterative | |
|---|---|---|---|---|---|---|---|---|
| | | SLiMFast | S-ERM | S-EM | Counts | ACCU | CATD | SSTF |
| Stocks | 0.1 | **0.856** | 0.780 | 0.691 | 0.628 | 0.754 | 0.724 | 0.754 |
| | 1 | **0.914** | 0.909 | 0.824 | 0.831 | 0.755 | 0.836 | 0.755 |
| | 5 | **0.925** | 0.875 | 0.897 | 0.849 | 0.760 | 0.897 | 0.753 |
| | 10 | **0.922** | 0.908 | 0.881 | 0.846 | 0.755 | 0.911 | 0.757 |
| | 20 | **0.922** | 0.912 | 0.918 | 0.817 | 0.765 | 0.915 | 0.757 |
| Demos | 0.1 | **0.732** | 0.694 | 0.663 | 0.525 | 0.554 | 0.511 | 0.678 |
| | 1 | **0.793** | 0.754 | 0.679 | 0.523 | 0.587 | 0.538 | 0.678 |
| | 5 | **0.826** | 0.804 | 0.813 | 0.555 | 0.587 | 0.538 | 0.679 |
| | 10 | 0.851 | **0.859** | 0.858 | 0.556 | 0.683 | 0.558 | 0.678 |
| | 20 | **0.900** | 0.899 | 0.897 | 0.503 | 0.708 | 0.564 | 0.679 |
| Crowd | 0.1 | 0.843 | 0.549 | 0.560 | 0.78 | **0.873** | 0.855 | 0.769 |
| | 1 | **0.926** | 0.872 | 0.837 | 0.911 | 0.914 | 0.888 | 0.781 |
| | 5 | **0.946** | 0.927 | 0.877 | 0.945 | **0.946** | 0.898 | 0.784 |
| | 10 | 0.945 | 0.931 | 0.914 | 0.954 | **0.956** | 0.935 | 0.783 |
| | 20 | 0.952 | 0.945 | 0.932 | **0.967** | 0.959 | 0.945 | 0.839 |
| Genomics | 0.1 | **0.567** | 0.542 | 0.543 | 0.557 | 0.546 | 0.532 | 0.556 |
| | 1 | **0.586** | 0.532 | 0.534 | 0.583 | 0.554 | 0.532 | 0.528 |
| | 5 | **0.613** | 0.534 | 0.536 | 0.586 | 0.559 | 0.549 | 0.537 |
| | 10 | **0.687** | 0.544 | 0.548 | 0.588 | 0.571 | 0.561 | 0.526 |
| | 20 | **0.720** | 0.563 | 0.571 | 0.5968 | 0.575 | 0.576 | 0.556 |

Panel B: Relative difference (%) between SLiMFast and other methods.

| | TD (%) | SLiMFast | S-ERM | S-EM | Counts | ACCU | CATD | SSTF |
|---|---|---|---|---|---|---|---|---|
| Average Accuracy across Datasets | 0.1 | **0.749** | -14.4 | -18.04 | -16.94 | -9.04 | -12.54 | -8.04 |
| | 1 | **0.805** | -4.72 | -10.71 | -11.52 | -12.71 | -13.20 | -14.81 |
| | 5 | **0.827** | -5.13 | -5.64 | -11.32 | -13.83 | -12.93 | -16.82 |
| | 10 | **0.851** | -4.78 | -5.99 | -13.53 | -12.92 | -12.92 | -19.41 |
| | 20 | **0.874** | -5.00 | -5.03 | -17.46 | -13.93 | -14.13 | -18.97 |

Comparing SLiMFast against methods that make strong independence assumptions, i.e., Counts, ACCU, and CATD, we see that SLiMFast yields accuracy improvements of more than 10% for almost all amounts of training data. In some cases, e.g., for Demonstrations, the absolute accuracy improvements are $\sim 0.3$—a relative improvement of more than 50%. This is because sources in the Demonstrations dataset are not independent as they correspond to online news media exhibiting correlations (Appendix D). In such cases, SLiMFast can effectively identify source correlations as it makes no assumptions over sources.

On the other hand, when sources are truly independent, as in Crowd, ACCU exhibits marginally better performance–especially for extremely small amounts of training data. This is due to the modeling assumptions of ACCU matching the way source observations were actually generated. For extremely sparse datasets, like Genomics, SLiMFast can yield accuracy improvements of up to 25% since domain-specific features allow it to recover source correlations more effectively than previous methods.

We now turn our attention to Sources-ERM and Sources-EM that use discriminative models. In most cases, SLiMFast outperforms both methods, thus, providing evidence that domain-specific features allow us to solve data fusion more accurately. SLiMFast yields an average accuracy improvement of 9.82% for small amounts of training data ($\leq 5\%$). For larger amounts of training data, i.e., 10% and 20%, all models yield comparable results.

**Table 3: Error for estimated source accuracies for varying training data. We focus on methods that use probabilistic semantics. Bold indicates the best performing method. Sources-ERM and Sources-EM are shown as S-ERM and S-EM.**

| Dataset | TD (%) | Discriminative | | | Generative | |
|---|---|---|---|---|---|---|
| | | SLiMFast | S-ERM | S-EM | Counts | ACCU |
| Stocks | 0.1 | **0.009** | 0.023 | 0.025 | 0.166 | 0.107 |
| | 1 | 0.01 | 0.008 | **0.006** | 0.021 | 0.108 |
| | 5 | 0.009 | **0.003** | **0.003** | 0.008 | 0.107 |
| | 10 | **0.002** | 0.008 | 0.004 | 0.012 | 0.096 |
| | 20 | 0.004 | **0.003** | 0.004 | 0.007 | 0.084 |
| Demos | 0.1 | **0.093** | **0.093** | **0.093** | 0.253 | 0.308 |
| | 1 | **0.093** | **0.093** | **0.093** | 0.218 | 0.298 |
| | 5 | **0.093** | 0.094 | **0.093** | 0.104 | 0.202 |
| | 10 | 0.094 | 0.094 | 0.094 | **0.074** | 0.183 |
| | 20 | 0.096 | 0.095 | 0.097 | **0.046** | 0.164 |
| Crowd | 0.1 | 0.027 | 0.102 | 0.112 | 0.292 | **0.013** |
| | 1 | **0.009** | 0.017 | 0.025 | 0.138 | 0.013 |
| | 5 | **0.008** | 0.009 | 0.016 | 0.051 | 0.012 |
| | 10 | **0.008** | 0.009 | 0.010 | 0.036 | 0.011 |
| | 20 | 0.008 | **0.007** | 0.009 | 0.031 | 0.009 |

Finally, we focus on SSTF which leverages semi-supervised learning to exploit the presence of labeled data. As shown in Table 2, SLiMFast always gives 13% more accurate results than SSTF on average. In certain cases, we observe accuracy improvements of more than 30%. We see that ACCU and CATD also outperform SSTF when altered to exploit the presence of labeled data.

*Takeaways.* Domain-specific features allow SLiMFast to obtain accurate data fusion results with a notably small amount of training data. In some cases ground truth data on only 1% of the objects allow SLiMFast to retrieve results that are more than 90% accurate. In datasets where source observations are not independent, such as Demonstrations, SLiMFast can identify the true value of objects more accurately (with absolute accuracy improvements of more than 0.3 in certain cases) than existing data fusion models. This is because, SLiMFast's discriminative model does not make any distributional assumptions over data sources, while existing data fusion approaches (e.g., ACCU and CATD) make strong independence assumptions for data sources. In general improvements, as the ones above, enable us to switch from data fusion outputs of moderate accuracy, i.e., 70%, to outputs that are 90% accurate.

### 5.2.2 Estimating the Accuracy of Data Sources

We now evaluate the ability of different data fusion methods to obtain low-error estimates of the true, unknown accuracy of data sources. We focus on models that follow probabilistic semantics, i.e., the trustworthiness of data sources is quantified via the notion of accuracy. The results are reported in Table 3.

For all configurations SLiMFast yields an average error for source accuracies less than 0.1. The error by other discriminative models (i.e., Sources-ERM and Sources-EM) are comparable to that of SLiMFast. On the other hand, the estimation-error obtained by generative Counts and ACCU is significantly higher for cases where either the available ground truth is very limited or the conditional independence assumption of their generative models does not hold.

*Takeaways.* In addition to estimating the true values of objects accurately, SLiMFast can also estimate the true accuracy of data sources with low-error. In fact, SLiMFast exhibits estimation-errors that are $2\times$ to $10\times$ lower than competing techniques.

**Table 4: Evaluating SLiMFast's optimizer at choosing between EM and ERM as we vary the amount of training data (TD). We report the accuracy score of SLiMFast-ERM and SLiMFast-EM, their relative difference and SLiMFast's optimizer decision. Bold entries indicate the best performing algorithm.**

| Dataset | TD (%) | Optimizer's Decision | Correct | Diff. (%) | SLiMFast ERM | SLiMFast EM |
|---|---|---|---|---|---|---|
| Stocks | 0.1 | EM | Y | 0.0 | **0.856** | **0.856** |
| | 1 | ERM | Y | 7.8 | **0.914** | 0.848 |
| | 5 | ERM | Y | 3.6 | **0.925** | 0.893 |
| | 10 | ERM | Y | 2.2 | **0.922** | 0.902 |
| | 20 | ERM | Y | 3.8 | **0.922** | 0.888 |
| Demos | 0.1 | EM | Y | 2.7 | 0.713 | **0.732** |
| | 1 | EM | Y | 6.6 | 0.744 | **0.793** |
| | 5 | EM | Y | 3.8 | 0.796 | **0.826** |
| | 10 | EM | Y | 0.0 | **0.851** | **0.851** |
| | 20 | EM | N | 2.3 | **0.921** | 0.900 |
| Crowd | 0.1 | EM | Y | 2.18 | 0.825 | **0.843** |
| | 1 | ERM | Y | 0.15 | **0.926** | 0.912 |
| | 5 | ERM | Y | 0.74 | **0.946** | 0.939 |
| | 10 | ERM | Y | 0.32 | **0.945** | 0.942 |
| | 20 | ERM | Y | 0.42 | **0.952** | 0.948 |
| Genomics | 0.1 | EM | Y | 12.1 | 0.506 | **0.567** |
| | 1 | EM | Y | 12.2 | 0.522 | **0.586** |
| | 5 | EM | Y | 4.4 | 0.587 | **0.613** |
| | 10 | EM | Y | 19.3 | 0.576 | **0.687** |
| | 20 | EM | Y | 8.9 | 0.661 | **0.720** |

*Omitted Comparison.* We omit CATD and SSTF from the comparisons since the former does not follow probabilistic semantics—source trustworthiness is measured via normalized weights across data sources—and the latter does not estimate the accuracy of sources. We also do not use Genomics as sources have a low number of observations, thus, their true accuracies cannot be estimated reliably.

### 5.2.3 Optimizer Evaluation

We measure the accuracy of SLiMFast-ERM, SLiMFast-EM at predicting the true values of objects and also report the algorithm that SLiMFast's optimizer chooses using the statistical model introduced in Section 4.3. For Algorithm 2 in the optimizer we set the threshold parameter $\tau$ to 0.1. The results are reported in Table 4. As shown, SLiMFast's optimizer can accurately choose the algorithm (EM or ERM) that yields the best data fusion output. While in most cases the accuracy scores of SLiMFast-EM and SLiMFast-ERM are comparable—the average accuracy improvement obtained by using the optimizer is 2.4%— there are cases where selecting correctly between EM and ERM yields a relative accuracy improvement of more than 7%, which in turn leads to results that are more than 90% accurate. Such accuracy improvements can be significant in sensitive applications where an accuracy of above 90% for the estimated true values of objects is required. The error for estimated source accuracies of SLiMFast-EM and SLiMFast-ERM is comparable and similar to that of SLiMFast (see Table 3).

Our experimental results indicate that the simple model in Section 4.3 is effective at analyzing heterogeneous datasets and correctly predicts the relative performance between expectation maximization (EM) and empirical risk minimization (ERM). For instance, we observe that SLiMFast's optimizer correctly selects to use the ERM algorithm for all amounts of ground truth in the Stocks dataset, where the density, i.e., the probability of a source providing an observation for an object, of this dataset is 0.99, and the average accuracy of data sources is below 0.5. In contrast, for Demonstrations and Genomics, SLiMFast's optimizer estimates a higher source accuracy and chooses to run EM, corresponding to the correct choice in almost all cases. Finally, for Crowd it can correctly identify the crossover point between EM and ERM as we increase the amount of available ground truth.

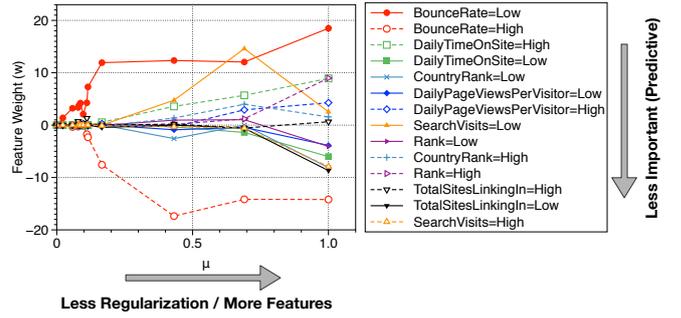

**Figure 6: Lasso path for features used in Stocks. Higher $x$-axis values correspond to lower regularization penalties.**

Finally, to evaluate the robustness of our optimizer, we vary the threshold parameter $\tau \in \{0.01, 0.1, 0.5, 1.0\}$. For Stock, our optimizer correctly selects to always run ERM for all values of $\tau$. Similarly for Genomics it always selects to run EM. For Crowd it makes no mistakes as it correctly switches to ERM after 1% of training data is revealed for all values of $\tau$. For Demonstrations and $\tau \geq 0.5$ our optimizer selects to run EM for $\leq 5\%$ of ground truth and ERM afterwards, thus, making no mistake. For $\tau = \{0.01\}$, our optimizer switches to ERM only for 40% of training data, hence, making two mistakes.

*Takeaways.* SLiMFast's optimizer can effectively choose between EM and ERM, thus, obviating the need for non-expert users to reason about which learning algorithm to use. Our experimental evaluation reveals that the simple model described in Section 4.3 can accurately detect which of learning algorithms, EM or ERM, will perform better for heterogeneous data fusion instances.

## 5.3 Studying Additional Functionalities

We now show how domain-specific features can be used in SLiM-Fast to obtain insights about the accuracy of sources and to estimate the quality of sources for which no observations are available.

### 5.3.1 Important Features of Source Accuracies

Users are often interested not only in how accurate a source is but the factors or features that affect its accuracy. We describe how domain-specific features in SLiMFast can be used to evaluate the accuracies of data sources. We describe how coupling SLiM-Fast with a standard statistical technique, called the *Lasso path*, both recovers insights about source accuracies stated in previous work and can be used to provide novel insights. *Lasso path* [37] is a standard technique to inspect the importance of features in discriminative probabilistic models. The idea is to examine how the feature-weights of an $L_1$-regularized model change as the regularization penalty varies: a high penalty means that fewer features obtain non-zero weights, thus, are used in the model, while a low regularization penalty allows more features to be used.

In lasso path, important features obtain non-zero weights for high regularization penalties and at the same time their absolute weight keeps increasing as the penalty decreases. Typically, the results of a lasso path are displayed as a plot; an example for the Stocks dataset is shown in Figure 6. On the x-axis we have a parameter $\mu \in [0, 1]$ that is inversely related to the regularization penalty and on the y-axis the weights of features in the model.

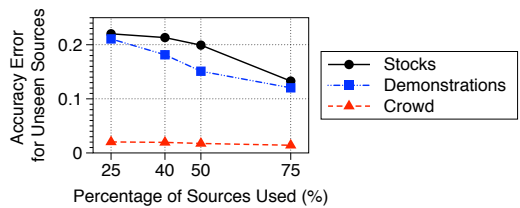

**Figure 7: Using features to estimate source accuracy.**

For Stocks, the most important features correspond to daily usage statistics, such as the "Bounce Rate" (with low bounce rate implying higher accuracy), and the "Daily Time on Site" (with low implying low accuracy). On the other hand, the number of "Total Sites Linking In", i.e., a proxy for PageRank, is not found to be important. This recovers a recent experimental result showing that PageRank is *not* correlated with the accuracy of web-sources [11]. Lasso path can be applied to any input provided to SLiMFast. In Appendix E, we show the Lasso path for the Crowd dataset, and we make the observation that the labor-channel via which workers are hired at CrowdFlower can be predictive of a worker's accuracy.

### 5.3.2 Source Quality Initialization

We examine how the output of SLiMFast, can be used to answer the problem of source quality initialization [25], i.e., the task of estimating the accuracy of newly available sources, from which no observations are available. The key idea is to just use the domain-specific features to predict the accuracy of sources. We use the learned weights of domain-specific features to predict the accuracy of new sources and measure the average absolute error between the estimated source accuracy and the true accuracy of sources.

We conduct the following experiment: for Stock, Demonstrations and Crowd, we restrict the number of sources given as input to SLiMFast, varying the percentage of used sources to be in $\{25\%, 40\%, 50\%, 75\%\}$. After learning the model parameters the feature weights are used to predict the accuracy of unseen sources. The results are shown in Figure 7. We see that the estimation error decreases as more sources are revealed to SLiMFast. For Stocks and Demonstrations, the error is larger compared to the case when we have access to the data of a source. For Crowd we can reliably predict the accuracy of unseen data sources even when only 25% of all sources are available.

## 6. RELATED WORK

The prior work related to this paper can be placed in a few categories; we describe each of them in turn:

**Data fusion.** There has been a significant amount of work on data fusion methods, including approaches that follow probabilistic semantics [9, 11, 30, 43], optimization-based techniques [22, 23], and iterative models [15, 29, 39]. All these methods only use the available source observations to estimate the trustworthiness of data sources and solve data fusion. On the other hand, SLiMFast can also use domain knowledge to solve data fusion more accurately.

**Quality guarantees for data fusion.** Several data fusion methods come with convergence guarantees [9] or confidence intervals for the estimated source trustworthiness [22]. Nevertheless, no existing method comes with guarantees on how close the estimated source accuracies are to the *true accuracies of data sources*. In contrast to them, SLiMFast comes with rigorous guarantees on its error rate for the estimated source accuracies, as well as its object truth estimates.

The only work prior to ours that comes with theoretical guarantees on the error rate of accuracy estimates is from the crowdsourcing community [8, 17, 20, 18] where data sources correspond to human workers. Moreover, recent work in crowdsourcing [16, 42] showed that EM-based Bayesian data fusion models can retrieve the true accuracies of data sources. Nevertheless, these results come with strong generative assumptions that only hold for restricted data fusion instances. Also, none of the proposed approaches consider exogenous features characterizing human workers. Our techniques might be of independent interest to the crowdsourcing community.

**Explanations for data fusion.** Understanding the output of data fusion is crucial for non-expert users [25]. Recent work [13] has considered the problem of providing explanations on the output of data fusion algorithms. The generated explanations correspond to compact summaries of the decisions made by the fusion algorithm during its execution. The summaries are also coupled with examples from the input source data to promote interpretability. In our work we follow an orthogonal direction. Instead of relying on source data and presenting the user with a trace of the fusion algorithm, we leverage the presence of domain specific features, and use techniques like L1-regularization [37] to present users with the most informative features that affect source accuracies.

**Efficiency of data fusion.** Most existing data fusion methods adopt iterative procedures like EM to estimate the quality of sources and compute the true values of objects [25]. This can be time consuming, especially when data fusion is applied on large scale data [24]. To address this challenge, recent literature has proposed the use of Map-Reduce based techniques [11] and has introduced streaming data fusion methods [44]. We show that when sufficient ground truth is available, we can avoid time consuming iterative algorithms entirely by using empirical risk minimization and efficient algorithms such as stochastic gradient descent.

## 7. CONCLUSION

We expressed data fusion as a learning and inference problem over discriminative probabilistic graphical models. This formulation enabled us to obtain data fusion models with rigorous guarantees and answer a series of open-problems in data fusion. We introduced SLiMFast, the first model that combines cross-source conflicts with domain-specific features for data fusion. Our theoretical and experimental analysis showed that domain-specific features not only lead to better source accuracy estimates but also allow us to identify the true value of objects more accurately. We also studied the tradeoff space between the quality of data fusion results when expectation maximization (EM)—a standard technique for learning the parameters of data fusion models—and empirical risk minimization (ERM) are used to learn SLiMFast's parameters. Extending this analysis to semi-supervised settings is an exciting future direction. We also proposed a simple model for estimating the performance of each algorithm and built an optimizer that automatically selects the best algorithm for learning the parameters of a given data fusion model. Our experiments confirmed the effectiveness of our optimizer for a variety of real-life setups. Finally, we showed that SLiMFast can be used to address open-problems in data fusion such as source reliability initialization.

# APPENDIX

## A. RADEMACHER COMPLEXITY

We provide a discussion on Rademacher complexity [26], one of the main tools used to derive the theoretical results presented in Section 4.2. The notion of Rademacher complexity [26] measures the richness of a class of functions and is used for excess risk bounds and generalization bounds. The Rademacher complexity of a set of functions $\mathcal{H}$ and a training set size $n$ is given by:

$$\mathcal{R}_n(\mathcal{H}) = (2/n) E_{\sigma_1,\ldots\sigma_n}[\sup_{h\in\mathcal{H}}(|\sum_{i=1}^{n} \sigma_i h(z_i)|)]$$

where $\sigma_i$ for $1 \leq i \leq n$ are independent random variables that take value $1$ or $-1$ with probability $1/2$ each, and $z_i$s are $n$ i.i.d training samples drawn from a distribution $\mathcal{D}$. We let $\mathcal{L}$ denote the set of $L_w$ for all $|K|$-dimensional vectors $w$. For our setting, $\mathcal{H}$ is simply $\mathcal{L}$, and the $z_i$'s are source-object pairs. $h(z_i)$ is simply $L_w(s, o)$. Since our loss is a function of a linear function of parameters $w$ and those of the source object pairs, we can use a well known bound on the Rademacher complexity of our class of loss functions. Specifically:

$$\mathcal{R}_n(\mathcal{L}) = O(\sqrt{|K|/n}\log(n)) \qquad (5)$$

Now we state an excess risk bound that uses Rademacher complexity. For any function class $\mathcal{H}$, for a set $S$ of $n$ i.i.d training samples $z_1, z_2, \ldots z_n$ from distribution $\mathcal{D}$, the empirical risk minimizer is defined as:

$$h_{erm} = \mathsf{argmin}_{h\in\mathcal{H}} \sum_{i=1}^{n} h(z_i)$$

Moreover, let the optimal function be

$$h_\infty = \mathsf{argmin}_{h\in\mathcal{H}} E_{z\sim\mathcal{D}}[h(z)]$$

Then with high probability (over the draws of $z_i$s), we have [26]:

$$E_{z\sim\mathcal{D}}[h_{erm}(z)] \leq E_{z\sim\mathcal{D}}[h_\infty(z)] + O(\mathcal{R}_n(\mathcal{H})) \quad (6)$$

Rademacher complexity also lets us quantify the rate of *uniform convergence*. Specifically, for any $h \in \mathcal{H}$, let

$$h(S) = (1/|S|)\sum_{z\in S} h(z) \text{ and } h_E = E_{z\sim\mathcal{D}}[h(z)]$$

When $S \sim \mathcal{D}^m$, we would like to upper bound $\max_{h\in\mathcal{H}} |h(S) - h_E|$ as a function of $m$. For any $\delta$ in $(0, 1)$, we have with probability $\geq 1 - \delta$,

$$\max_{h\in\mathcal{H}} |h(S) - h_E| \leq 2\mathcal{R}_n(\mathcal{H}) + \sqrt{\log(1/\delta)/2n} \quad (7)$$

## B. PROOFS

We provide the proofs for the theorems in Section 4.2.

### B.1 Proof of Theorem 1

This theorem follows from the uniform convergence result in Equation 7. For any constant $\delta$, and $|G| = n$, the $\sqrt{\log(1/\delta)/2n}$ term is subsumed by the $\mathcal{R}_n(\mathcal{H})$ term, because for our case,

$$\mathcal{R}_n(\mathcal{H}) = O(\sqrt{|K|/n}\log(n))$$

due to Equation 5. Thus we simply have, for any $h' \in \mathcal{H}$,

$$|h'(S) - h'_E| \leq \max_{h\in\mathcal{H}} |h(S) - h_E| \leq O(\sqrt{|K|/n}\log(n))$$

Replacing $h(S)$ by $L_G(w)$ and $h_E$ by $L(w)$ gives us our theorem.

### B.2 Proof of Theorem 2

In our setting, $E_{z\sim\mathcal{D}}[h_{erm}(z)]$ from Equation 6 is simply $L_w$, while $E_{z\sim\mathcal{D}}[h_\infty(z)]$ is $L_{w^*}$, and $\mathcal{H}$ is the family of losses for all weight vectors. Then applying Equation 5 to the $\mathcal{R}_n(\mathcal{H})$ term gives us Theorem 2.

### B.3 Proof of Theorem 3

We consider $|G| = n$. We consider two cases based on the probability $p$ of a source observing an object. If $p = \Omega(\log(n)/\delta)$, then with high probability, the majority value of observations on each object equals the true value of the object. In this case, we can just compute the majority and treat it as ground truth, and apply Theorem 2. The number of labeled source-object pairs in this case is $O(|S||O|p)$, substituting $n = |S||O|p$ that value gives us our required result. The second case is where $p = O(\log(n)/\delta)$. In this case we drop objects that have $< 2$ observations, and for other objects, randomly drop all but two observations. This reduces the problem to a slightly different problem, which is easier to solve. We now solve the reduced problem in the rest of the section.

The reduced problem is as follows: We have $n = |O|$ objects. For each object, we choose a pair of sources uniformly at random, and both the chosen sources make an observation on the object. In this case, we show that we can estimate source accuracies such that $\frac{1}{|S|}\sum_{s\in S} D_{KL}(A_s || A_s^*) \leq O(\frac{\log|O|}{|S|\delta} + \sqrt{\frac{|K|}{|O||S|}}\log(|O||S|))$. This, along with $p = O(\sqrt{n}/\delta)$, will prove our theorem.

For each object 0, we randomly designate one of its two observing sources as 'primary', denoted $S(o)$. For any source $s$, le $O_s$ denote the set of objects such that $S(o) = s$. For any object $o$, let Agree($o$) equal 1 if the two sources observing $o$ agreed on it, and 0 otherwise. Our accuracy estimation algorithm has three steps:

1. Estimate $A_E = \sum_{s\in S}(2A_s^* - 1)$. Let $A'_E$ denote our estimate.

2. For each source $s$, compute $a_s = (\sum_{o\in O_s}(2|S|\text{Agree}(o) - (|S| - A'_E)))/2A'_E$.

3. Choose $w$ to minimize $\sum_{s\in S} a_s \log(\text{logistic}(w\cdot F_s)) + (|O_s| - a_s)\log(1 - \text{logistic}(w \cdot F_s))$.

Then we prove that the chosen $w$ from the third step must be such that accuracy estimates $A_s = \text{logistic}(w \cdot F_s)$ satisfy our guarantee.

**Step 1:** We first show that step 1 can be carried out, such that our estimate $A'_E$ has a relative error of $O(1/|S|\delta + 1/\sqrt{n})$, with high probability. Relative error of $\epsilon$ here means that $A'_E$ is in interval $((1 - \epsilon)A_E, (1 + \epsilon)A_E)$.

To begin with, note that $2A_s^* - 1 \in [\delta, 1 - \delta]$ for all $s$. For any $o \in O$, the expected value $E[\text{Agree}(o)]$ is

$$(1/|S|(|S|-1))\sum_{s_1,s_2\in S, s_1\neq s_2} A_{s_1}^* A_{s_2}^* + (1 - A_{s_1}^*)(1 - A_{s_2}^*)$$

since the pair of sources is chosen randomly, and sources $s_1$ and $s_2$ agree on an object if they are both correct or both wrong. Moreover, $A_{s_1}^* A_{s_2}^* + (1 - A_{s_1}^*)(1 - A_{s_2}^*)$ equals $1/2 + (2A_{s_1}^* - 1)(2A_{s_2}^* - 1)/2$. Thus expected value of $2|S|(|S|-1)\text{Agree}(o)$ is

$$2\sum_{s_1,s_2\in S, s_1\neq s_2} A_{s_1}^* A_{s_2}^* + (1 - A_{s_1}^*)(1 - A_{s_2}^*)$$

$$=\sum_{s_1,s_2\in S, s_1\neq s_2} 1 + (2A_{s_1}^* - 1)(2A_{s_2}^* - 1)$$

$$=\sum_{s_1,s_2\in S} 1 + (2A_{s_1}^* - 1)(2A_{s_2}^* - 1) - \sum_{s\in S} 1 + (2A_s^* - 1)^2$$

$$=|S|(|S|-1) + \sum_{s_1,s_2\in S}(2A_{s_1}^* - 1)(2A_{s_2}^* - 1) - \sum_{s\in S}(2A_s^* - 1)^2$$

$$=|S|(|S|-1) + \left(\sum_{s\in S}(2A_s^* - 1)\right)^2 - \sum_{s\in S}(2A_s^* - 1)^2$$

$$=|S|(|S|-1) + A_E^2 - \sum_{s\in S}(2A_s^* - 1)^2$$

Thus expected value of $|S|(|S|-1)(2\text{Agree}(o)-1)$ equals $A_E^2 - \sum_{s\in S}(2A_s^* - 1)^2$. Moreover, $A_E \geq |S|\delta$ and $2A_s^* - 1 \leq 1$ for all $s$, so

$$A_E^2 = \sum_{s\in S}(2A_s^* - 1)A_E \geq \sum_{s\in S}(2A_s^* - 1)|S|\delta$$

$$\geq \sum_{s\in S}(2A_s^* - 1)|S|\delta(2A_s^* - 1) = \sum_{s\in S}(2A_s^* - 1)^2|S|\delta$$

This gives us $A_E^2 \geq A_E^2 - \sum_{s\in S}(2A_s^* - 1)^2 \geq A_E^2(1 - 1/(|S|\delta))$, so we can estimate $A_E$ by computing

$$|S|(|S|-1)/|O|(\sum_{o\in O} 2\text{Agree}(o) - 1)$$

and taking it's square root. Since the sum over $o \in O$ involves i.i.d variables, we can use the Chernoff bound, showing that the sum, and hence estimate of $A_E$, has a relative error of $O(1/(|S|\delta) + 1/\sqrt{n})$. Let $A'_E$ denote our noisy estimate of $A_E$ obtained above. This completes step 1.

**Step 2, 3:** Step 2 is simple, and is used to compute the loss function in Step 3. We now show that the $w$'s computed in step 3 give accurate estimates. For that, we need to define two loss functions, $L_1$ and $L_2$. $L_1$ will be the loss function we actually care about (log-loss of source accuracies). However, we cannot directly estimate $L_1$ without ground truth. So we will define another loss function $L_2$, which will be shown to be close to $L_1$, and which can be estimated from samples of source conflicts.

First, we define $L_1(w) = (1/|S|) \sum_{s \in S} A_s^* \log(\text{logistic}(w \cdot F_s)) + (1 - A_s^*) \log(1 - \text{logistic}(w \cdot F_s))$. The weights $w^*$ that minimize this loss function are precisely the weights that give us $A_s^* = \text{logistic}(w^* \cdot F_s)$ for all $s$. Unfortunately, we cannot empirically estimate this loss directly.

In step 3, we minimize the 'empirical' loss function $L_{emp}(w)$ given by $\sum_{s \in S} a_s \log(\text{logistic}(w \cdot F_s)) + (|O_s| - a_s) \log(1 - \text{logistic}(w \cdot F_s))$.

$L_{emp}(w)$ can also be written as $\sum_{o \in O} a_o \log(\text{logistic}(w \cdot F_{S(o)})) + (1 - a_o) \log(1 - \text{logistic}(w \cdot F_{S(o)}))$ where

$$a_o = (2|S|\text{Agree}(o) - (|S| - A_E'))/2A_E'$$

We now define a loss $L_2$ such that the $L_{emp}(w)$ is a 'sampled' version of $L_2$. Specifically, define $L_2(w)$ to be:

$$E[a_o \log(\text{logistic}(w \cdot F_{S(o)})) + (1 - a_o) \log(1 - \text{logistic}(w \cdot F_{S(o)}))]$$

where the expectation is taken over objects (random choices over the sources that observe the object, and over the observations of those sources). For any object $o \in O_s$, we define $B_s^*$ such that

$$B_s^* = (1/(2|S| - 2))(|S| - 1 + (2A_s^* - 1)(A_E - 2A_s^* + 1))$$

Thus, we have $E[\text{Agree}(o)] = B_s^*$. So

$$E_{o \in O_s}[a_o] = (2|S|B_s^* - (|S| - A_E'))/2A_E'$$

Moreover, we now have

$L_2(w) =$
$(1/|S|)(\sum_{s \in S}(2|S|B_s^* - (|S| - A_E'))) \log(\text{logistic}(w \cdot F_s))/2A_E'$
$+ (1 - (2|S|B_s^* - (|S| - A_E'))/2A_E') \log(1 - \text{logistic}(w \cdot F_s)))$

Our proof now proceeds in two steps. (a) We show that $L_1$ and $L_2$ are very close, upto scaling (b) We show that the weight vector $w$ chosen to minimize $L_{emp}(w)$, also achieves a very low value for $L_2$, using standard Rademacher complexity bounds.

We show (b) first, since it is straightforward. We choose $w_e$ to minimize $L_{emp}$, i.e. $w_e = \arg\min_w \sum_{o \in O} a_o \log(\text{logistic}(w \cdot F_{S(o)})) + (1 - a_o) \log(1 - \text{logistic}(w \cdot F_{S(o)}))$. Now $L_2$ is simply the expected value of this quantity. Let $\mathcal{L}$ denote the family of $L_2(w)$ over all $w$. Then, since each member of the family consists of a function of a linear function of $w$, the Rademacher complexity $\mathcal{R}_n(\mathcal{L}) = O(\sqrt{k/n} \log(n))$. Thus,

$$L_2(w_e) \leq L_2(w') + O(\sqrt{k/n} \log(n))$$

where $w'$ is the weight vector that minimizes $L_2(w)$.

Now we show (a). To begin with, we will show that $B_s^*$ is very close to $(1/2|S|)(|S| + (2A_s^* - 1)(A_E))$. We have $B_s^* = (1/(2|S| - 2))(|S| - 1 + (2A_s^* - 1)(A_E - 2A_s^* + 1))$, so

$|B_s^* - (1/2|S|)(|S| + (2A_s^* - 1)(A_E))|$
$= 1/(2|S|(|S|-1))|(|S|(|S|-1) + |S|(2A_s^* - 1)(A_E - 2A_s^* + 1))$
$\quad - |S|(|S|-1) + (|S|-1)(2A_s^* - 1)(A_E))|$
$= 1/(2|S|(|S|-1))(2A_s^* - 1)||S|(A_E - 2A_s^* + 1) - (|S|-1)A_E)|$
$= 1/(2|S|(|S|-1))(2A_s^* - 1)|A_E - |S|(2A_s^* - 1)|$
$\leq 1/(2|S|(|S|-1))(2A_s^* - 1)|S|$
$\leq 1/(2|S| - 2)$

Since $B_s^* \geq 1/2$, it means that $B = (1/2|S|)(|S| + (2A_s^* - 1)(A_E))$ estimates all $B_s^*$ with a relative error of $\leq 1/(|S| - 1)$.

Moreover, we saw earlier that $A_E'$ estimates $A_E$ with relative error of $O(1/(|S|\delta) + 1/\sqrt{n})$. Also note that if we replace $B_s^*$ with $B$ and $A_E'$ with $A_E$ in $L_2$, then it would become equal to $L_1$:

$(2|S|B - (|S| - A_E))/2A_E$
$=2|S|((1/2|S|)(|S| + (2A_s^* - 1)(A_E)) - (|S| - A_E))/2A_E$
$=((|S| + (2A_s^* - 1)(A_E)) - (|S| - A_E))/2A_E$
$=(|S| + 2A_s^*A_E - A_E - |S| + A_E)/2A_E$
$=(2A_s^*A_E)/2A_E = A_s^*$

Similarly, the other term (corresponding to $1 - a_o$) becomes $1 - A_s^*$. With this insight, we can show that $L_1$ and $L_2$ are close. The numerator, as seen above, was equal to $2A_s^*A_E \geq 2(1/2)|S|\delta$ (since $A_E \geq |S|\delta$). The multiplicative error in the $2|S|B$ term is $1/(|S| - 1)$, thus absolute error is $O(1)$ since $B < 1$. This will cause a relative error of $O(1/(|S|\delta))$ in numerator $2A_s^*A_E$. Similarly, relative error in $A_E'$ is $1/(|S|\delta) + 1/\sqrt{n}$, which causes a relative error of $1/(|S|\delta) + 1/\sqrt{n}$ in $2A_s^*A_E$. Thus the total relative error in $2A_s^*A_E$ is at most $O(1/(|S|\delta) + 1/\sqrt{n})$. We can show a similar result for the second terms of $L_1$ and $L_2$. Thus $L_2$ approximates $L_1$ with a relative error of $O(1/(|S|\delta) + 1/\sqrt{n})$. Moreover, any $w$ we consider for learning with $n$ samples will satisfy $\log(\text{logistic}(w \cdot F_s)) = O(\log(n))$. Thus for any such $w$, $|L_1(w) - L_2(w)| = O(\log(n)/(|S|\delta) + \log(n)/\sqrt{n})$.

Now, let $w^*$ be the weights that give us true accuracies i.e. $A_s^* = \text{logistic}(w \cdot F_s)$ and, $w_e$ is the weight chosen by our algorithm:

$L_1(w_e) \leq L_2(w_e) + O(\log(n)/(|S|\delta) + \log(n)/\sqrt{n})$
$\leq L_2(w') + O(\sqrt{k/n} \log(n)) + O(\log(n)/(|S|\delta) + \log(n)/\sqrt{n})$
$\leq L_2(w^*) + O(\sqrt{k/n} \log(n)) + O(\log(n)/(|S|\delta) + \log(n)/\sqrt{n})$
$\leq L_1(w^*) + O(\sqrt{k/n} \log(n)) + O(\log(n)/(|S|\delta) + \log(n)/\sqrt{n})$

The first and last step uses the closeness of $L_1$ and $L_2$. The second step uses the Rademacher complexity excess risk bound given earlier. The thid step is because $w'$ is defined as the $w$ that minimizes $L_2$. This shows that our chosen $w_e$ will $L_1$ loss value which is not much worse than that of the best weights $w^*$.

Moreover, $L_1(w) - L_1(w^*)$ is simply the average KL divergence with true accuracies. i.e.

$$L_1(w_e) - L_1(w^*) = (1/|S|) \sum_{s \in S} KL(A_s || A_s^*)$$

where $A_s = \text{logistic}(w_e \cdot F_s)$. This gives us

$$(1/|S|) \sum_{s \in S} KL(A_s || A_s^*) = O(\log(n)/(|S|\delta) + \sqrt{k/n} \log(n))$$

This proves our claim, and using $p = O(\log(n)/\delta)$ finishes our proof for the theorem.

## C. RUNTIME ANALYSIS

We measure the total wall-clock runtime of each data fusion methods for all datasets in Table 5. Reported runtimes correspond to end-to-end execution with data pre-processing and loading. For DeepDive-based methods, this includes loading data into a database, compiling the input data to a factor graph and then running inference and learning. For python methods, pre-processing corresponds to only loading input data from raw files.

We compare SLiMFast with Sources-ERM and Sources-EM. We see that: (i) when SLiMFast's optimizer switched from EM to ERM, SLiMFast's runtime is reduced significantly (see Crowd for TD =

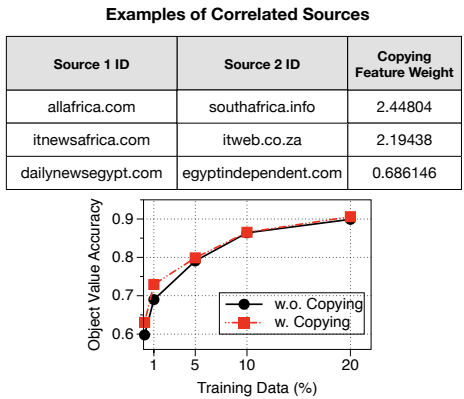

Figure 8: Detecting source copying for Demonstrations.

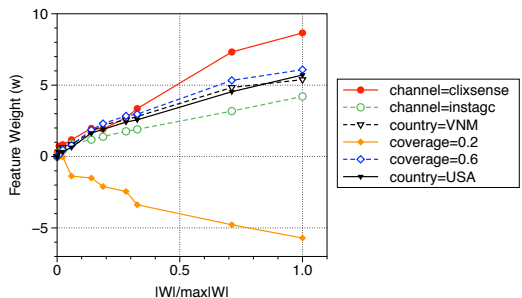

Figure 9: Lasso path for features used in Crowd. Higher $x$-axis values correspond to lower regularization penalties.

Table 5: Wall-clock runtimes (in seconds) for data fusion methods. We report end-to-end execution including data loading for each method.

| Dataset | TD (%) | Discriminative | | | Generative | | Iterative | |
|---|---|---|---|---|---|---|---|---|
| | | SLiMFast | S-ERM | S-EM | Counts | ACCU | CATD | SSTF |
| Stocks | 0.1 | 60.71 | 40.12 | 102.53 | 2.77 | 5.04 | 7.04 | 150.39 |
| | 1 | 62.51 | 41.91 | 108.97 | 3.03 | 5.70 | 5.48 | 174.31 |
| | 5 | 64.51 | 41.76 | 99.32 | 3.3 | 6.4 | 6.21 | 168.92 |
| | 10 | 68.51 | 40.82 | 105.67 | 3.59 | 6.4 | 6.96 | 156.31 |
| | 20 | 68.41 | 44.51 | 104.94 | 4.97 | 6.3 | 6.66 | 144.40 |
| Demos | 0.1 | 170.13 | 41.46 | 94.45 | 3.01 | 6.61 | 9.05 | 109.39 |
| | 1 | 180.42 | 41.76 | 109.76 | 2.32 | 6.29 | 10.01 | 102.75 |
| | 5 | 168.06 | 47.63 | 98.62 | 2.27 | 5.66 | 9.11 | 104.94 |
| | 10 | 171.39 | 62.26 | 106.25 | 2.51 | 5.71 | 10.18 | 101.74 |
| | 20 | 189.23 | 75.90 | 128.69 | 2.41 | 5.10 | 9.18 | 120.59 |
| Crowd | 0.1 | 104.03 | 46.41 | 83.61 | 4.06 | 6.41 | 6.22 | 37.31 |
| | 1 | 56.45 | 50.92 | 88.79 | 3.57 | 5.72 | 5.53 | 37.21 |
| | 5 | 66.22 | 50.38 | 109.74 | 3.33 | 7.18 | 5.84 | 37.55 |
| | 10 | 66.13 | 50.08 | 89.47 | 2.57 | 6.14 | 5.39 | 38.08 |
| | 20 | 78.9 | 51.44 | 89.35 | 3.44 | 6.18 | 6.92 | 37.36 |
| Genomics | 0.1 | 67.60 | 60.07 | 67.76 | 1.11 | 7.01 | 29.41 | 6.94 |
| | 1 | 65.48 | 62.52 | 65.70 | 1.14 | 7.17 | 29.03 | 7.04 |
| | 5 | 65.66 | 61.63 | 66.65 | 1.26 | 6.33 | 27.42 | 7.05 |
| | 10 | 65.63 | 63.48 | 65.71 | 1.35 | 6.97 | 28.34 | 6.04 |
| | 20 | 65.61 | 61.16 | 65.93 | 1.15 | 6.78 | 28.14 | 6.46 |

0.1% versus TD= 1%), (ii) incorporating domain-specific features does not incur drastic runtime changes as SLiMFast's runtime is comparable to that of Sources-ERM and Sources-EM depending on the learning algorithm used. We see that in most cases, the end-to-end runtime is around a minute. For Demonstrations we observe increased runtimes due since SLiMFast's optimizer selects to run EM for most cases. Nonetheless, the overall runtime is still around 3 minutes. While SLiMFast's runtime is higher than that of baseline methods the accuracy improvements obtained by using SLiMFast (up to 44% as shown in Table 2) justify the use of SLiMFast.

Table 6: End-to-end v.s. learning-and-inference-only runtime (in seconds) for DeepDive-based methods on Genomics.

| TD (%) | End-to-end | | | Learning and Inference Only | | |
|---|---|---|---|---|---|---|
| | SLiMFast | S-ERM | S-EM | SLiMFast | S-ERM | S-EM |
| 0.1 | 67.60 | 60.07 | 67.76 | 8.87 | 6.63 | 8.91 |
| 1 | 65.48 | 62.52 | 65.70 | 8.79 | 6.73 | 8.83 |
| 5 | 65.66 | 61.63 | 66.65 | 8.79 | 6.49 | 8.78 |
| 10 | 65.63 | 63.48 | 65.71 | 8.70 | 6.78 | 8.90 |
| 20 | 65.61 | 61.16 | 65.93 | 8.75 | 6.78 | 8.79 |

Comparing SLiMFast with Python-based models we observe significant differences in the end-to-end runtime. To understand if this difference is due to data loading or due to the learning and inference algorithm used by SLiMFast, we compared the end-to-end runtime of SLiMFast, Sources-ERM, and Sources-EM against their learning-and-inference-only runtime. We report the results for Genomics in Table 6. As shown, most of the time is spent in compiling the input data to a factor graph and the time spent in learning and inference, i.e., solving data fusion, is comparable to python-based methods. Similar results were observed for all datasets.

## D. COPYING SOURCES

We present how more complex data fusion methods can be implemented in SLiMFast. We examine data fusion methods that identify data sources that copy from each other when solving data fusion. Source copying can be modeled via the following intuition: if two sources make the same mistakes they have a higher probability of copying from each other [9].

To model this in SLiMFast, we extend its probabilistic model with a set of Boolean features for all source pairs, such that each feature takes the value "True" only when the two sources in it agree on their observations for an object $o \in O$ but variable $v_o$ is set to a value different than the one reported by the sources. SLiMFast's probabilistic model remains a logistic regression model.

To evaluate how well the above model captures source copying, we use the Demonstrations dataset for which sources correspond to online news portals, thus, copying is expected to occur. We compare the accuracy for object values obtained by SLiMFast when copying is modeled and when it is not. For simplicity, no domain-specific features were used. The results are shown in Figure 8. For very small amounts of training data modeling copying leads to better performance. We also provide examples of sources that are found to be copying from each other in Demonstrations. As shown, sources providing news for the same region (e.g., Egypt) or the same categories (e.g., business) exhibit high weights for the features modeling copying.

## E. LASSO PATH FOR CROWD

Figure 9 shows the lasso path plot for features used in Crowd. We show the paths for the first two features from channel, city and coverage that activated. The first feature that activates is channel "clixsense". Also it is interesting to observe that "coverage=0.6" (relatively high coverage) has a high positive weight while "coverage=0.2" (low coverage) has a high negative weight.